\documentclass[12pt]{article}

\begin{document}

\title{Dendrites and conformal symmetry}

\author{ Juan M. Romero$^{(1)}$\thanks{jromero@correo.cua.uam.mx}, Carlos Trenado$^{(2)}$\thanks{trenado@cdb-unit.de}\\
[0.5cm]\\
\it $^{(1)}$Departamento de Matem\'aticas Aplicadas y Sistemas\\
\it Universidad Aut\'onoma Metropolitana-Cuajimalpa\\
\it M\'exico, D.F   05300, M\'exico\\
\it$^{(2)}$ Deparment of Neurology, Freiburg University\\
\it Breisacherstrasse 64, 79106\\
\it Freiburg im Breisgau, Germany\\
[0.3cm]}

\date{}

\pagestyle{plain}

\maketitle

\begin{abstract}

Progress toward characterization of structural and biophysical properties of neural dendrites together with recent findings emphasizing their role in neural computation, has propelled growing interest in refining existing theoretical models of electrical propagation in dendrites while advocating novel analytic tools. 
In this paper we focus on the cable equation describing electric propagation in dendrites with different geometry.
When the geometry is cylindrical we show that the cable equation is invariant under the Schr\"odinger group and by using the dendrite parameters, a representation of the 
Schr\"odinger  algebra is provided. Furthermore, when the geometry profile is parabolic we show that the cable equation is equivalent to the Schr\"odinger equation for the 1-dimensional free particle, which is invariant under the Schr\"odinger group. Moreover, we show that there is a family of dendrite geometries for which the cable equation is equivalent to the Schr\"odinger equation for the 1-dimensional conformal quantum mechanics.

\end{abstract}

\section{Introduction}

Dendrites are neuron branched projections first discovered by Ramon y Cajal at the end of the 19th century \cite{ramon:gnus}. It is well-documented that dendrites arborize  through a volume of brain tissue so as to collect information in the form of synaptic inputs. A single neuron may receive as many as 200,000 synaptic inputs trough its dendrites, which account for up to 99\% of a neuron's membrane \cite{kandel:gnus}.\\

With respect to dendrite's morphology, a trade off between the metabolic costs of dendrite elaboration and the need to cover the receptive field have been suggested as basic principles to determine their size and shape, while the connectivity between neurons has been implicated in shaping the geometry and spatial orientation of dendritic arborization \cite{jan:gnus,fiala:gnus}. In this regard, novel experimental techniques enable the characterization and identification of molecules involved in different aspects of dendritic development \cite{scott:gnus}. \\

Focusing on neural processing, theoretical studies have emphasized the role of dendrites in altering the potential range of single neuron computations while being considered as sub-units of integration with sigmoidal or Heaviside activation functions \cite{mel:gnus,gurney:gnus,poirazia:gnus,poirazib:gnus,caze:gnus}. As such, these dendritic sub-units greatly expand the computational capacity of a neuron as in the case of 
the feature storage capacity \cite{poirazia:gnus}, the computation of binocular disparity \cite{archie:gnus} and the computation of object-feature binding problems \cite{legenstein:gnus}. Also, recent experimental studies provide evidence on the role of dendritic excitability as an essential component of behaviourally relevant computations in neurons \cite{spencer:gnus}. On the basis of these observations, a deeper understanding and refinement of theoretical models dealing with electrical propagation in dendrites is crucial. \\

The first model addressing the electrical conductivity of dendrites via the passive cable theory was formulated by Rall \cite{ralla:gnus,rallb:gnus}. Such a formulation remains relevant till today because passive properties of dendritic membranes provide the essential steps in the process of filtration and integration carried out by dendrites. Within this framework, a dendrite can be described as a cable with circular cross-section and diameter $d(x).$ Now, if $V(x,t)$ denotes the electrical voltage in the dendrite, the following equation is satisfied \cite{ermentrout:gnus} 
\begin{eqnarray}
c_{M}\frac{\partial V(x,t) }{\partial t}= \frac{1 }{4r_{ L} d(x) } \frac{\partial  }{\partial x} \left( d^{2}(x) \frac{\partial V(x,t) }{\partial x}\right)-  \frac{V(x,t) }{r_{ M}},
\label{eq:c1}
 \end{eqnarray}
where $c_{M}$ denotes the specific membrane capacitance, $r_{M}$ represents the membrane resistance and $r_{L}$ denotes the longitudinal resistance. In studying electric propagation in dendrites by equation (\ref{eq:c1}) it is commonly assumed that their diameter is constant. However, consideration of dendrites with varying diameters is exemplified by different type of dendrite structural specializations e.g. sites of synaptic contact, also including dendrites which present varicosities as in the case of dendrites in retina amacrine cells, the cerebellar dentate nucleus and the lateral vestibular nucleus presenting as well as 
cortical pyramidal and olfactory bulb cells presenting bulbous enlargements on dendritic tips \cite{kandel:gnus,fiala:gnus}.\\

 With regard to novel mathematical techniques dealing with electric propagation in dendritic trees, the formalism of path integrals  commonly derived in quantum physics has been highlighted \cite{abbot:gnus}. Then, it might be possible  that techniques developed in quantum physics can  be employed to study  electrical propagation in a dendrite. For example, quantum mathematical techniques are useful to solve differential  equations.  In fact,   it is well known that symmetry methods in quantum physics enable defining the properties of a system without the need to solve all the equations involved in its description.  In spite of this, symmetry methods have barely been applied to study differential equations that describe biological systems. It is worth mentioning that an important symmetry in physics  is given by the   conformal symmetry.  This symmetry appears in systems as non-relativistic free particle \cite{hagen:gnus,hagen1:gnus}, atomic models \cite{jackiw3:gnus}, quark models \cite{brodsky2:gnus}, black-holes \cite{jackiw4:gnus}  and string theory \cite{maldacena2:gnus}. Below we show that this symmetry appears in the equation (\ref{eq:c1}) for different dendrite geometries.\\ 

In this paper we consider a more realistic family of dendrites that includes the types mentioned previously. Specifically, we consider dendrites with diameter
\begin{eqnarray}
d(x)= d_{0}(1+a x)^{\nu}, \label{eq:cable-family}
\end{eqnarray}
where $\nu,d_{0},$ and $a$ are real constants. Note that different dendritic geometries can be obtained by changing the parameter $\nu$, for example if $\nu=0,$ the dendrite geometry is cylindrical, while if $\nu=1$ the dendrite geometry is conical. First, we show  that when $\nu=0$ the cable equation (\ref{eq:c1}) is invariant under the Schr\"odinger group, which is  a conformal group, and  using the dendrite parameters a representation of the   Schr\"odinger  algebra is obtained.  Second, we show that when $\nu=2$  the 
cable equation is equivalent to the Schr\"odinger equation for the 1-dimensional free particle, which is invariant under the Schr\"odinger group.  
Third, we study the cable equation when $\nu \not =0 $ and $\nu\not = 2,$ in this case we show that the cable equation  is equivalent to the Schr\"odinger equation for   the 1-dimensional conformal quantum mechanics.  In addition, in this last case,   using the dendrite parameters a representation of the conformal algebra is obtained.\\

This paper is organized as follows: in section 2 we provide a brief overview of the free particle and conformal quantum mechanics. In section 3, we study the particular case of cylindrical dendrites. In section 4, we address the case of dendrites with parabolic profile. In section 5, we study the relationship between dendrites and conformal quantum mechanics. In section 6, a summary is provided.  

\section{Non-relativistic conformal symmetry}

In this section we provide a brief overview about the one dimensional non-relativistic free particle and the conformal quantum mechanics.

\subsection{Free particle}

The action for the $1$-dimensional free non-relativistic particle is given by
\begin{eqnarray}
S=\int dt \frac{m}{2}\left(\frac{dz}{dt}\right)^{2},\label{eq:action}
\end{eqnarray}
where $m$ is the particle mass. Note that if we take the conformal transformations
\begin{eqnarray}
t^{\prime}=\frac{\alpha t+\beta}{\gamma t+\delta},\qquad z^{\prime}=\frac{lz+vt+c}{\gamma t+\delta}, \qquad l^{2}=\alpha \delta -\beta \gamma\not = 0, 
\label{eq:conformes}
\end{eqnarray}
where $\alpha, \beta, \gamma,\delta, l, v, c$ are constants, then the action (\ref{eq:action}) is transformed as 
\begin{eqnarray}
S^{\prime}=\int dt^{\prime} \frac{m}{2}\left(\frac{dz^{\prime}}{dt^{\prime}}\right)^{2}=S+\frac{m}{2}
\int dt \left(\frac{d \phi(z,t)}{dt}\right), \label{eq:action-t}
\end{eqnarray}
where 
\begin{eqnarray}
 \phi(z,t)= \frac{1}{l^{2}}\left(   2lv z+ v^{2}t-\frac{\gamma \left(  lz +vt +c \right)^{2}}{ \gamma t+\delta }  \right).\label{eq:phace}
\end{eqnarray}
Then, the free non-relativistic particle dynamics is invariant under the conformal transformations (\ref{eq:conformes}). Such  transformations (\ref{eq:conformes}) include temporal and  spatial translations, Galileo's transformations, anisotropic scaling and the special conformal transformations. \\

In quantum mechanics,  the wave function for the one dimensional non-relativistic free particle satisfies the Schr\"odinger equation 
\begin{eqnarray}
 i\hbar \frac{\partial\Psi\left(z,t\right)}{\partial t} =
 -\frac{\hbar^{2}}{2m} \frac{\partial ^{2} \Psi (z,t) }{\partial z^{2}}. \label{eq:schrodinger}
\end{eqnarray}
This  equation  is invariant under the conformal coordinate transformations (\ref{eq:conformes}), where the wave function transforms as
\begin{eqnarray}
\Psi^{\prime}\left(z^{\prime},t^{\prime}\right)=\left( \sqrt{ \gamma t +\delta }\right) e^{\frac{im}{2\hbar}\phi(z,t) }\Psi(z,t), \label{eq:onda-conforme}
\end{eqnarray}
and $\phi(z,t)$ is given by  (\ref{eq:phace}).\\

Furthermore, the following operators 
\begin{eqnarray}
\hat P&=&-i\hbar \frac{\partial }{\partial z},\label{eq:opsch1} \\
\hat H&=&\frac{\hat P^{2}}{2m},\label{eq:hamiltonianf}\\
\hat G&=&t\hat P-mz,\\
\hat K_{1}&=& t\hat H-\frac{1}{4}\left( z\hat P+\hat P z\right),\\
\hat K_{2}&=& t^{2}\hat H-\frac{t}{2}\left(  z\hat P+\hat Pz\right)+\frac{m}{2}z^{2}.\label{eq:opsch2}
\end{eqnarray}
are generators of the transformations (\ref{eq:conformes}). These  operators satisfy the Schr\"odinger algebra 
\begin{eqnarray}
\left[\hat P,\hat H\right]&=&0,\label{eq:sch1tt} \\
\left[\hat P,\hat K_{1}\right]&=&\frac{i\hbar }{2} \hat P,\\
\left[\hat P,\hat K_{2}\right]&=&i \hbar \hat G,\\
\left[\hat P,\hat G\right]&=&i\hbar m,\\
\left[\hat H,\hat K_{1}\right]&=&i\hbar \hat H,\\
\left[\hat H,\hat G\right]&=&i\hbar \hat P,\\
\left[\hat H,\hat K_{2}\right]&=&2i\hbar \hat K_{1},\\
\left[\hat K_{1},\hat K_{2}\right]&=&i\hbar \hat K_{2},\\
\left[\hat K_{1},\hat G\right]&=&\frac{i\hbar }{2}\hat G,\\
\left[\hat K_{2},\hat G\right]&=&0.\label{eq:sch2tt}
\end{eqnarray}
It is possible to show that the operators (\ref{eq:sch1tt})-(\ref{eq:sch2tt}) are conserved. This algebra is the so-called Schr\"odinger algebra.
The conformal symmetry for the free Schr\"odinger equation was found by 
Niederer and Hagen in 1972 \cite{hagen1:gnus,hagen:gnus}.\\ 

In what follows we show that the conformal symmetry and the Schr\"odinger algebra are present in the cable equation when the dendrite geometry is cylindrical or parabolic.

\subsection{Conformal quantum mechanics}

An interesting system in quantum mechanics is given by the so-called conformal quantum mechanics  \cite{jackiw6:gnus}. This system appears in different contexts, from black-holes to atomic physics and quark models \cite{jackiw2:gnus,jackiw3:gnus,jackiw4:gnus}.\\

The Schr\"odinger equation for the $1$-dimensional conformal quantum mecha\-nics is 
\begin{eqnarray}
i\hbar \frac{\partial \psi(x,t)}{\partial t} =\hat H\psi(x,t), \qquad \hat H=-\frac{\hbar^{2}}{2m} \frac{\partial^{2} }{\partial x^{2}} +\frac{g}{x^{2}} .\label{eq:cqmh}
\end{eqnarray}
Here $g$ denotes a coupling constant. When $g\not =0$ the Galileo's and translation symmetries are broken. Then, this system is not invariant under all conformal transformations (\ref{eq:conformes}). In fact, the conformal quantum mechanics is only invariant under the coordinate transformations
\begin{eqnarray}
t^{\prime}=\frac{\alpha t+\beta}{\gamma t+\delta},\qquad x^{\prime}=\frac{lx}{\gamma t+\delta}, \quad l^{2}=\alpha\delta-\beta \gamma\not =0.   
\label{eq:conformes-2}
\end{eqnarray}
In this case,  in order to keep the wave equation invariant under the transformations (\ref{eq:conformes-2}), the wave function must be transformed as

\begin{eqnarray}
 \psi^{\prime}\left(x^{\prime},t^{\prime}\right)= \sqrt{ \gamma t +\delta } e^{ i\frac{m}{2\hbar} \Phi(x,t) } \psi(x,t)
\label{eq:conformes-3}
\end{eqnarray}
where
\begin{eqnarray}
\label{eq:funcion}
 \Phi(x,t)= -\frac{\gamma x^{2}}{ \gamma t+\delta } .
\end{eqnarray}
Moreover, the generators are given by $\hat H$ and 
\begin{eqnarray}
\hat K_{1}&=& t\hat H-\frac{1}{2}\left( x\hat P-\frac{i}{2}\right),\\
\hat K_{2}&=&t^{2}\hat H-t\left(x\hat P-\frac{i}{2}\right)+\frac{m x^{2}}{2}
\end{eqnarray}
and the algebra 
\begin{eqnarray}
[ \hat H, \hat K_{1}]=i\hbar  \hat H, \qquad [\hat H,\hat K_{2}]=2i\hbar  \hat K_{1},\qquad [\hat K_{1},\hat K_{2}]=i\hbar \hat K_{2}
\end{eqnarray}
is satisfied. Using this algebra, it is possible to show that the operator $\hat H,\hat K_{1},\hat K_{2}$ is conserved.\\
 
Now, we show that the Schr\"odinger equation for the conformal quantum mechanics (\ref{eq:cqmh}) is equivalent to the cable equation (\ref{eq:c1}) for different dendrite geometries.

\section{Cylindrical Case}
 
If a dendrite has cylindrical geometry, the diameter is given by  $d(x)=d_{0}=$constant. In this case the cable equation (\ref{eq:c1}) becomes
\begin{eqnarray}
c_{M}\frac{\partial V (x,t) }{\partial t}= \frac{d_{0} }{4r_{ L} } \frac{\partial^{2} V(x,t) }{\partial x^{2}}-  \frac{V (x,t) }{r_{ M}}.
\label{eq:cableusual0} 
\end{eqnarray}
The solution for this equation can be found in \cite{ermentrout:gnus}. \\

Notice that if we take 
\begin{eqnarray}
 V_{cil} (x,t)= e^{-\frac{t}{c_{M}r_{M} }}\psi(x,t), \label{eq:p-c}
 \end{eqnarray}
we arrive to
\begin{eqnarray}
\frac{\partial \psi(x,t) }{\partial t}= \frac{d_{0}}{4r_{ L} c_{M}} \frac{\partial^{2} \psi(x,t) }{\partial x^{2}}.
\label{eq:cableusual}
 \end{eqnarray}
This  equation is a Schr\"odinger-like equation (\ref{eq:schrodinger}). Then, due that the Schr\"odinger equation (\ref{eq:schrodinger})  
is invariant under the conformal transformations (\ref{eq:conformes}), the 
equation (\ref{eq:cableusual}) is invariant under the same transformations.  In this case the function $\psi(x,t)$ becomes  
\begin{eqnarray}
\psi^{\prime}\left(x^{\prime},t^{\prime}\right)=\left( \sqrt{\gamma t+\delta}\right)
e^{-\frac{c_{M}r_{L}}{ d_{0} }\phi(x,t) }\psi(x,t),
\end{eqnarray}
where $\phi(x,t)$ is given by (\ref{eq:phace}). Then, the cable equation (\ref{eq:cableusual0}) is invariant under conformal transformations (\ref{eq:conformes})
where $V(x,t)$ is transformed as  
\begin{eqnarray}
V^{\prime}\left(x^{\prime},t^{\prime}\right)=\left( \sqrt{\gamma t+\delta}\right)e^{\frac{-\gamma t^{2}+(\alpha-\delta)t +\beta}{\delta t+\delta} }
e^{-\frac{c_{M}r_{L}}{ d_{0}  }\phi(x,t) }V(x,t),
\end{eqnarray}
here  $\phi(x,t)$ is given by  (\ref{eq:phace}). \\

In addition, note that the equation (\ref{eq:cableusual0}) can be written as
\begin{eqnarray}
-\frac{\partial V(x,t) }{\partial t}= \hat HV(x,t), 
\end{eqnarray}
where 
\begin{eqnarray}
\hat H= \frac{d_{0} }{4r_{ L}c_{M} } \hat P^{2}+ \frac{1 }{c_{M}r_{ M}}, \qquad \hat P=-i\frac{\partial }{\partial x}. \label{eq:c-cil}
\end{eqnarray}
We can see  that this operator is the Hamiltonian for the non-relativistic free particle (\ref{eq:hamiltonianf}) with "mass" 
\begin{eqnarray}
m=\frac{2r_{L}c_{M}}{d_{0}}.\label{eq:efmc}
\end{eqnarray}
Furthermore, the following operators 
\begin{eqnarray}
\hat P&=&-i\frac{\partial }{\partial x},\label{eq:opsch1} \\
\hat H_{0}&=&   \frac{d_{0} }{4r_{ L}c_{M} }  P^{2},\label{eq:hamiltonian}\\
\hat G&=&t\hat P- \frac{2r_{L}c_{M}}{d_{0}}x,\\
\hat K_{1}&=& t\hat H-\frac{1}{4}\left( x\hat P+\hat P x\right),\\
\hat K_{2}&=& t^{2}\hat H-\frac{t}{2}\left(  x\hat P+\hat Px\right)+\frac{r_{L}c_{M}}{d_{0}}x^{2}.\label{eq:opsch2}
\end{eqnarray}
are generators of the transformations (\ref{eq:conformes}). It can be shown  that the   operators (\ref{eq:c-cil}) and (\ref{eq:opsch1})-(\ref{eq:opsch2}) satisfy the Schr\"odinger algebra 
\begin{eqnarray}
\left[\hat P,\hat H_{0}\right]&=&0,\label{eq:sch1} \\
\left[\hat P,\hat K_{1}\right]&=&\frac{i}{2} \hat P,\\
\left[\hat P,\hat K_{2}\right]&=&i  \hat G,\\
\left[\hat P,\hat G\right]&=&i  \frac{2r_{L}c_{M}}{d_{0}} ,\\
\left[\hat H_{0},\hat K_{1}\right]&=&i \hat H_{0},\\
\left[\hat H_{0},\hat G\right]&=&i\hat P,\\
\left[\hat H_{0},\hat K_{2}\right]&=&2i \hat K_{1},\\
\left[\hat K_{1},\hat K_{2}\right]&=&i \hat K_{2},\\
\left[\hat K_{1},\hat G\right]&=&\frac{i }{2}\hat G,\\
\left[\hat K_{2},\hat G\right]&=&0.\label{eq:sch2}
\end{eqnarray}
Then, when the dendrite geometry is cylindrical, the cable equation (\ref{eq:c1}) is equivalent to the Schr\"odinger equation for the non-relativistic free particle with "mass" given by (\ref{eq:efmc}). For this reason the conformal symmetry is present in this kind of dendrite geometry. 

\section{Parabolic case}

If the dendrite geometry is parabolic, we have   
\begin{eqnarray}
d(x)= d_{0}\left( 1+ax\right)^{2}, \label{eq:diametre}
\end{eqnarray}
where $d_{0}$ and $a$ are constants. In this case the cable equation (\ref{eq:c1}) can be written as
\begin{eqnarray}
c_{M}\frac{\partial V(x,t) }{\partial t}= \frac{d_{0} \left( 1+ax\right)^{2} }{4r_{ L} } \frac{\partial^{2} V(x,t) }{\partial x^{2} }  + \frac{ad_{0} \left( 1+ax\right) }{r_{ L} } \frac{\partial V(x,t) }{\partial x } -  \frac{V(x,t) }{r_{ M}}. \label{eq:cable2}
 \end{eqnarray}
Now, using the change of variable 
\begin{eqnarray}
 1+ax=e^{z} \label{eq:changeov}, 
 \end{eqnarray}
in the equation (\ref{eq:cable2}),  we obtain 
\begin{eqnarray}
\frac{\partial V(z,t) }{\partial t}= \frac{a^{2} d_{0} }{4c_{M} r_{ L} } \frac{\partial^{2} V(z,t) }{\partial z^{2} }  + \frac{3a^{2} d_{0}}{4c_{M}r_{ L} } \frac{\partial V(z,t) }{\partial z } -  
\frac{V(z,t) }{c_{M}r_{ M}}. \label{eq:cablecv}
 \end{eqnarray}
In addition, if we take 
\begin{eqnarray}
V(z,t)=  e^{\left( -\frac{3z}{2}-\left( \frac{9a^{2}d_{0} }{16r_{ L} c_{M} }  +  \frac{1}{r_{ M} c_{M}  } \right) t\right) } \psi(z,t), \label{eq:propose}
 \end{eqnarray}
the equation (\ref{eq:cablecv}) becomes 
\begin{eqnarray}
\frac{\partial \psi(z,t) }{\partial t}= \frac{a^{2} d_{0} }{4c_{M} r_{ L} } \frac{\partial^{2} \psi (z,t) }{\partial z^{2} } ,  \label{eq:cabletrans}
 \end{eqnarray}
which  is a Schr\"odinger-like equation (\ref{eq:schrodinger}). \\

\subsection{Symmetries}

Now, due  that the equation (\ref{eq:cabletrans}) is Schr\"odinger-like  equation  (\ref{eq:schrodinger}), it  is invariant under the conformal transformations.  In fact, using the change of variable  (\ref{eq:changeov}), the coordinate  transformations  
(\ref{eq:conformes}) can be written  as 
\begin{eqnarray}
x^{\prime}= \frac{1}{a} \left( \left( 1+ax\right)^{\left(\frac{l}{\gamma t +\delta}\right)} e^{\left( \frac{vt+c}{\gamma t +\delta }\right)}-1\right) ,\qquad 
t^{\prime}=\frac{\alpha t+\beta}{\gamma t +\delta}.
\end{eqnarray}
Through  a long but straightforward calculation, it can be shown  the cable equation (\ref{eq:cable2}) is invariant under the earlier transformations, where $V(x,t)$ becomes
\begin{eqnarray}
V^{\prime}\left(x^{\prime}, t^{\prime}\right)=\left( \sqrt{ \gamma t +\delta }\right)\left[ \left( 1+ax\right)^{ -\left( \frac{3}{2}\frac{l }{\gamma t+\delta}+\frac{4c_{M} r_{L} v }{a^{2} d_{0} l}\right)}   \right] e^{\Phi(x,t)} V\left(x, t\right),   
\end{eqnarray}
here 
\begin{eqnarray}
\Phi(x,t)&=&-\Bigg[\frac{3}{2}\frac{vt+c}{\gamma t+\delta } + \left( \frac{ 2c_{M} r_{L} v^{2}  }{a^{2}d_{0}l^{2} } +  \frac{9d_{0} a^{2}}{16 r_{ L} c_{M}}  +\frac{1}{r_{M} c_{M} } \right)t+  \nonumber\\
& & -\frac{2c_{M} r_{ L} \gamma }{a^{2} d_{0} l^{2}\left( \gamma t+\delta\right) }   \left(  l\ln (1+ax) +vt +c \right)^{2}   \Bigg].  
\end{eqnarray}

\subsection{ Schr\"odinger algebra}

The cable equation (\ref{eq:cable2}) can be written as 
\begin{eqnarray}
-\frac{\partial V(x,t) }{\partial t}= \hat H V(x,t),
 \end{eqnarray}
where 
\begin{eqnarray}
\hat H= \frac{d^{2}_{0} \left( 1+ax\right)^{2} }{4r_{ L} c_{M}} \hat P^{2}  -i \frac{a d_{0} \left( 1+ax\right) }{r_{ L} c_{M}} \hat P+  \frac{1}{r_{ M}c_{M}}, \qquad \hat P=-i\frac{ \partial}{\partial x}.
\label{eq:hamcable}
 \end{eqnarray}
Now, notice that using the operator 
\begin{eqnarray}
\hat \Pi = \frac{ \left( 1+ax\right)}{a} \hat P  -i \frac{3}{2}, \label{eq:pi}
 \end{eqnarray}
the Hamiltonian (\ref{eq:hamcable})  can be written as
\begin{eqnarray}
\hat H= \frac{ a^{2}d_{0}  }{4r_{ L} c_{M}} \hat \Pi^{2}  + \frac{9}{16 r_{L} c_{M} }+  \frac{1}{r_{ M}c_{M}},
 \end{eqnarray}
which is equivalent to the Hamiltonian for the non-relativistic free particle (\ref{eq:hamiltonianf}) with "mass" 
\begin{eqnarray}
\tilde m=\frac{2r_{L}c_{M}}{d_{0}a^{2}}.\label{eq:efm}
\end{eqnarray}
In addition we can construct the operators
\begin{eqnarray}
\hat H_{0}&=& \frac{ a^{2} d_{0}}{4r_{ L} c_{M}}\hat  \Pi^{2},\label{eq:1} \\
\hat G&=&t \hat \Pi -\frac{2r_{ L} c_{M}}{a^{2}} \ln(1+ax),\\
\hat K_{1}&=& t\hat H_{0} -\frac{1}{2} \left(\ln(1+ax)\hat \Pi+ \hat \Pi \ln(1+ax)   \right),\\
\hat K_{2}&=&t^{2} \hat H_{0} -\frac{t}{2} \left(  \ln(1+ax) \hat  \Pi+\hat \Pi   \ln(1+ax) +\frac{r_{L} c_{M}}{a^{2}}  [  \ln(1+ax) ]^{2} \right).\label{eq:4}
\end{eqnarray}
These operators are equivalent to the operators (\ref{eq:opsch1})-(\ref{eq:opsch2}). In fact, using the relationship  
\begin{eqnarray}
\left[ \ln(1 +ax),\hat \Pi\right] =i,
 \end{eqnarray}
it is possible to show that the operators (\ref{eq:pi}) and (\ref{eq:1})-(\ref{eq:4}) satisfy the Schr\"odinger algebra
\begin{eqnarray}
\left[\hat \Pi,\hat H_{0}\right]&=&0,\label{eq:sch1} \\
\left[\hat \Pi,\hat K_{1}\right]&=&\frac{i }{2} \hat \Pi,\\
\left[\hat \Pi,\hat K_{2}\right]&=&i  \hat G,\\
\left[\hat \Pi,\hat G\right]&=&i\frac{2c_{M}  r_{ L} }{a^{2}d_{0}}  ,\\
\left[\hat H_{0},\hat K_{1}\right]&=&i \hat H_{0},\\
\left[\hat H_{0},\hat G\right]&=&i \hat \Pi,\\
\left[\hat H_{0},\hat K_{2}\right]&=&2i \hat K_{1},\\
\left[\hat K_{1},\hat K_{2}\right]&=&i \hat K_{2},\\
\left[\hat K_{1},\hat G\right]&=&\frac{i}{2}\hat G,\\
\left[\hat K_{2},\hat G\right]&=&0,\label{eq:sch2}.
\end{eqnarray}
Thus, when the dendrite geometry is parabolical, the cable equation (\ref{eq:cable2}) can bee seen as a free Schr\"odinger equation (\ref{eq:schrodinger}) which a particular  change of coordinates.

\section{Dendrites and conformal quantum mechanics} 
 
In this section we study the cable equation when the dendrite diameter is given by (\ref{eq:cable-family}).
For this case the cable equation (\ref{eq:c1}) is 
\begin{eqnarray}
\frac{\partial V(x,t) }{\partial t}&=& \frac{d_{0} }{4r_{ L} c_{M}  } \left(1+ax\right)^{\nu} \frac{\partial^{2} V(x,t) }{\partial x^{2}}+ \frac{d_{0} \nu a}{2r_{ L} c_{M}  } \left(1+ax\right)^{\nu-1} \frac{\partial V(x,t) }{\partial x} \nonumber\\
& & -  \frac{V(x,t) }{r_{ M} c_{M}}.\label{eq:dgc}
 \end{eqnarray}
Now, using the change of variable 
\begin{eqnarray}
z=\frac{\left(1+ax\right)^{1-\frac{\nu}{2}}}{a\left(1-\frac{\nu}{2}\right)}, \label{eq:changeofv}
\end{eqnarray}
we arrive to
\begin{eqnarray}
\frac{\partial V(z,t) }{\partial t}= \frac{d_{0} }{4r_{ L} c_{M}  }\frac{\partial^{2} V(z,t) }{\partial z^{2}}+ \frac{3d_{0} \nu }{8r_{ L} c_{M}  \left(1-\frac{\nu}{2} \right)}  \frac{1}{z}\frac{\partial V(z,t) }{\partial z}  -  \frac{V(z,t) }{r_{ M}c_{M}}.
 \end{eqnarray}
Notice that when  $\nu =2$ or $a=0$ the change of variable (\ref{eq:changeofv}) is singular, these cases were previously studied.\\

Now, if we take 
\begin{eqnarray}
V(z,t) =\left(z\right)^{-\frac{3\nu }{4\left( 1-\frac{\nu}{2} \right)  }} e^{-\frac{t}{r_{M}c_{M}}}\psi(z,t),\label{eq:gcc}
\end{eqnarray}
the following equation 
\begin{eqnarray}
\frac{\partial \psi(z,t) }{\partial t}= \frac{d_{0} }{4r_{ L} c_{M}  }\frac{\partial^{2} \psi(z,t) }{\partial z^{2}}+ \frac{3d_{0} \nu (4-5\nu) }{16r_{ L} c_{M}  \left(2-\nu \right)^{2}} \frac{1}{z^{2}}\psi(z,t) 
\label{eq:dendrite-conformal}
 \end{eqnarray}
is obtained. In addition, if we take 
\begin{eqnarray}
\hat  H= -\frac{1  }{2m_{0} } \frac{\partial^{2}  }{\partial z^{2} } +\frac{g}{z^{2}}, \qquad m_{0}=\frac{2r_{ L} c_{M}}{d_{0}} ,\qquad g=\frac{3d_{0}\nu (5\nu -4) }{16r_{L} c_{M} (2-\nu)^{2}},\label{eq:cable-hamiltonian}
 \end{eqnarray}
the equation (\ref{eq:dendrite-conformal}) can be written as  
\begin{eqnarray}
-\frac{\partial \psi(z,t) }{\partial t}= \hat H \psi(z,t). 
\label{eq:dendrite-conformal1}
 \end{eqnarray}
Notice that the operator (\ref{eq:cable-hamiltonian})  is the Hamiltonian for the conformal quantum mechanics (\ref{eq:cqmh}).\\

Note that  the family of  dendrite diameters  (\ref{eq:cable-family}) is associated with the family of conformal Hamiltonians  (\ref{eq:hamiltonian}).  
Furthermore, observe that  for each Hamiltonian  (\ref{eq:hamiltonian}) we have 
two dendrite diameters, namely  each  Hamiltonian is associated with two $\nu$ values. For example,  $\nu=0$ represents a cylindrical dendrite  while $\nu=\frac{4}{5}$ represents a particular conical dendrite, but both give rise to the equation
\begin{eqnarray}
\frac{\partial \psi(z,t) }{\partial t}= \frac{d_{0} }{4r_{ L} c_{M}  }\frac{\partial^{2} \psi(z,t) }{\partial z^{2}}.
\end{eqnarray}
However, notice that the electrical voltage is not the same, as can be seen in the equation (\ref{eq:gcc}). 

\subsection{Symmetry}

As the  equation (\ref{eq:dendrite-conformal}) is equivalent to the Schr\"odinger equation for the conformal quantum mechanics (\ref{eq:cqmh}), then the equation (\ref{eq:dendrite-conformal}) is invariant under the coordinate transformations 
(\ref{eq:conformes-2}). In this case  
the function $\psi(z,t)$ transforms as 
\begin{eqnarray}
 \psi^{\prime}\left(z^{\prime},t^{\prime}\right)= \sqrt{ \gamma t +\delta } e^{ -\frac{d_{0}}{r_{L} c_{M}} \Phi(z,t) } \psi(z,t).
\label{eq:conformes-3}
\end{eqnarray}
Hence, the cable equation (\ref{eq:dgc}) is invariant under the coordinate transformations (\ref{eq:conformes-2}) where the voltage potential transforms as   
\begin{eqnarray}
V^{\prime}(x^{\prime},t^{\prime}) =\frac{\left(\gamma t +\delta \right)^{\frac{1+\nu }{2-\nu}}}{ a^{\frac{3\nu }{2\left( 2-\nu \right)  }}}  e^{-\frac{1}{r_{M}c_{M}} \left(\frac{-\gamma t^{2} +(\alpha -\delta)t +\beta }{ \gamma t +\delta }\right) } e^{\frac{4\gamma d_{0}}{a^{2} r_{M} c_{M}}\frac{\left(1+ax\right)^{2-\nu}} { (\gamma t+\delta) \left(2-\nu\right)^{2} } } V(x,t).
\end{eqnarray}

\subsection{Conformal algebra}

The equation (\ref{eq:dgc}) can be written as
\begin{eqnarray}
-\frac{\partial V(x,t)}{\partial t}=\hat {\bf H}V(x,t), 
 \end{eqnarray}
 where
\begin{eqnarray}
\hat {\bf H}=  -\frac{d_{0} }{4r_{ L} c_{M}  } \left(1+ax\right)^{\nu} \frac{\partial^{2}  }{\partial x^{2}}-\frac{d_{0} \nu a}{2r_{ L} c_{M}  } \left(1+ax\right)^{\nu-1} \frac{\partial  }{\partial x} 
+\frac{1 }{r_{ M} c_{M}}. \label{eq:H}
 \end{eqnarray}
Now, using the operator 
\begin{eqnarray}
\hat {\bf \Pi}=-i  \left( 1+ax\right)^{\frac{\nu}{2} } \frac{\partial }{\partial x }  - i\frac{3\nu a }{4} \left( 1+ax\right)^{\frac{\nu}{2} -1},
 \end{eqnarray}
the Hamiltonian (\ref{eq:H}) can be written as 
\begin{eqnarray}
\hat {\bf H}= \frac{ d_{0} }{4r_{ L} c_{M}}\hat {\bf \Pi}^{2}    + \frac{3a^{2} d_{0}  }{64 r_{ L} c_{M}} (5\nu-4) \left(1+ax\right)^{\nu-2} +\frac{1}{r_{ M}c_{M}}. 
 \end{eqnarray}
In addition the following relation
\begin{eqnarray}
\left[ \frac{\left(1+ax\right)^{1-\frac{\nu}{2}}}{a\left(1-\frac{\nu}{2}\right)}  ,\hat {\bf  \Pi} \right]= i
 \end{eqnarray}
is satisfied.\\
  
Then, the following operators 
\begin{eqnarray}
{\bf H}_{0}&=& \frac{ d_{0} }{4r_{ L} c_{M}}\Pi^{2}    + \frac{3a^{2} d_{0}  }{64 r_{ L} c_{M}} (5\nu-4) \left(1+ax\right)^{\nu-2},\\ 
\hat {\bf K}_{1}&=& t\hat H_{0}-\frac{1}{2}\left(\frac{\left(1+ax\right)^{1-\frac{\nu}{2}}}{a\left(1-\frac{\nu}{2}\right)}   \hat \Pi-\frac{i}{2}\right),\\
\hat {\bf K}_{2}&=&t^{2}\hat H_{0}-t\left(\frac{\left(1+ax\right)^{1-\frac{\nu}{2}}}{a\left(1-\frac{\nu}{2}\right)}  \hat \Pi -\frac{i}{2}\right)+ \frac{r_{L} c_{M}}{d_{0} } \frac{\left(1+ax\right)^{2-\nu}}{a^{2} \left(1-\frac{\nu}{2}\right)^{2} }
\end{eqnarray}
satisfy  the conformal algebra
\begin{eqnarray}
[ \hat {\bf H}_{0}, \hat {\bf K}_{1}]=i\hat {\bf H}_{0}, \qquad [\hat {\bf H}_{0},\hat {\bf K}_{2}]=2i\hat {\bf K}_{1},\qquad [\hat {\bf K}_{1},\hat {\bf K}_{2}]=i\hat {\bf K}_{2}.
\end{eqnarray}

\subsection{Solution}

In addition, if  we take
\begin{eqnarray}
\psi(z,t)=e^{-E t} \phi(z) 
 \end{eqnarray}
the equation (\ref{eq:dendrite-conformal}) becomes 
\begin{eqnarray}
E\phi(z) =\hat H\phi(z), \label{eq:dendrite-conforme}
 \end{eqnarray}
The solution for the equation (\ref{eq:dendrite-conforme}) is given by 
\begin{eqnarray}
\phi(z) =z^{\frac{1}{2}} \left[ AJ_{\sqrt{\frac{1}{4}+ \frac{3\nu(5\nu-4 )}{ 4(2-\nu)^{2} } }} \left( \sqrt{\frac{4c_{M}r_{L} E}{d_{0}}} z\right)+BN_{\sqrt{\frac{1}{4}+ \frac{3\nu(5\nu-4 )}{ 4(2-\nu)^{2} } }}\left( \sqrt{\frac{4c_{M}r_{L} E}{d_{0}}} z\right)\right], \nonumber
 \end{eqnarray}
where $J_{\alpha}(z)$ is a Bessel function, $N_{\alpha}(z)$ is a Neumann function and $A,B$ are constants. Then, the solution for the equation (\ref{eq:dgc}) is given by
\begin{eqnarray}
V(x,t) &=&\left(1+ax\right)^{\frac{(1+\nu) }{2}} e^{-\left(E+\frac{1}{c_{M}r_{ L}} \right)t}   \Bigg[ AJ_{\sqrt{\frac{1}{4}+ \frac{3\nu(5\nu-4 )}{ 4(2-\nu)^{2} } }}
 \left( \sqrt{\frac{4c_{M}r_{L} E}{d_{0}}}     \frac{\left(1+ax\right)^{\nu-\frac{1}{2}}}{a\left(\frac{\nu}{2}-1\right)}    \right)\nonumber\\
 & & +BN_{\sqrt{\frac{1}{4}+ \frac{3\nu(5\nu-4 )}{ 4(2-\nu)^{2} } }}\left( \sqrt{\frac{4c_{M}r_{L} E}{d_{0}}}   \frac{\left(1+ax\right)^{\nu-\frac{1}{2}}}{a\left(\frac{\nu}{2}-1\right)} \right)
\Bigg].
 \end{eqnarray}
The conformal quantum mechanics appears  in different contexts, from black-holes to atomic physics and quark models \cite{jackiw2:gnus,jackiw3:gnus,jackiw4:gnus}. Notably, this system describes a electrical voltage in different dendrite geometries.

\section{Summary}

In this paper we studied  the cable equation and its symmetries for different dendritic geometries. 
When the geometry is cylindrical we showed  that the cable equation is invariant under the Schr\"odinger  group,   and  using the dendrite parameters a representation of the   Schr\"odinger  algebra was  obtained. Furthermore, we showed that  when the  geometry is parabolical the cable equation is equivalent to the Schr\"odinger equation for the 1-dimensional free particle, which is invariant under the Schr\"odinger group.  In addition, it was  shown  that  the cable equation is equivalent to the Schr\"odinger equation for the 1-dimensional conformal quantum mechanics for a family of dendritic geometries. A generalized solution for the voltage of the considered family of dendrites is provided.   
From the neuroscience perspective, the considered family of dendrites include geometries that are more realistic when considering electric propagation in dendritic synapses as well as bulbous enlargements and varicosities that are present in various types of dendrites in the human brain.

\end{document}